\documentclass[twocolumn, showpacs]{revtex4}

\usepackage{graphicx}

\usepackage{dcolumn}

\usepackage{bm}

\begin{document}

\title{High-mobility AlAs quantum wells with out-of-plane valley occupation}

\date{\today}

\author{K.\ Vakili}

\author{Y.P.\ Shkolnikov}

\author{E. Tutuc}

\author{E.P.\ De Poortere}

\author{M.\ Padmanabhan}

\author{M.\ Shayegan}

\affiliation{Department of Electrical Engineering, Princeton
University, Princeton, NJ 08544}

\begin{abstract}

Employing state-of-the-art molecular beam epitaxy techniques to
grow thin, modulation-doped AlAs quantum wells, we have achieved a
low temperature mobility of 5.5 m$^2$/Vs with out-of-plane
occupation, an order of magnitude improvement over previous
studies. However, due to the narrow well width, mobilities are
still limited by scattering due to interface roughness disorder.
We demonstrate the successful implementation of a novel technique
utilizing thermally-induced, biaxial, tensile strain that forces
electrons to occupy the out-of-plane valley in thicker quantum
wells, reducing interface roughness scattering and allowing us to
achieve mobilities as high as 8.8 m$^2$/Vs.

\end{abstract}


\maketitle

Two-dimensional electron systems (2DESs) confined to AlAs quantum
wells (QWs) have revealed a variety of novel phenomena in recent
years. This has been made possible, in large part, by advances in
semiconductor growth techniques that have dramatically improved
the quality of these systems \cite{lay93, depoortere02}, allowing
the observation of behaviors that had previously been obscured by
disorder effects. These improvements have, however, only been
implemented in thick AlAs QWs in which electrons occupy the
valleys oriented in the QW plane. Recent work has demonstrated the
value of studying thinner AlAs QWs with the out-of-plane valley
occupied \cite{vakili04}, and there is therefore a desire to
reduce the obfuscating effects of disorder in these systems as
well.

Bulk AlAs has an indirect band gap with conduction band minima at
the six equivalent X-points of the Brillouin zone.  The Fermi
surface for electrons consists of three full (six half) prolate
ellipsoids, or valleys, with major axes parallel to each of the
crystal axes (Fig. 1). The conduction mass, in units of the bare
electron mass, is {\it m}$_{l}$ = 1.04 along each ellipsoid's
major (longitudinal) axis and {\it m}$_{t}$ = 0.21 along the minor
(transverse) axes \cite{momose99}. Confining the charges along a
crystal direction breaks the bulk valley degeneracy as a
consequence of the valleys' mass anisotropy. Our QWs are grown on
(001) GaAs substrates, and the valley oriented along this axis
(which we will heretofore refer to as the Z valley) is lower in
confinement energy with respect to the two in-plane valleys (X,Y)
due to its larger mass along [001]. There is a competing effect,
however, due to the lattice mismatch between AlAs and the GaAs
substrate that induces a biaxial, compressive strain, lowering the
X and Y valleys in energy with respect to the Z valley by
approximately 23 meV \cite{vankesteren89, maezawa92,
vandestadt96}.  Consequently, there is a crossover between Z and
(X,Y) valley occupation at a particular QW thickness, {\it
t}$_{c}$ $\sim$ 55 $\AA$, with thinner wells favoring the Z valley
and thicker wells favoring the X and Y.

The main difference between electrons occupying the Z or (X,Y)
valleys is the shape of the Fermi contour projected in the 2DES
plane. For the Z valley, the contour is isotropic, while the X and
Y each have anisotropic contours. Though such anisotropy can be
interesting, the Z valley's isotropic profile provides a simpler
system for the study of general properties of 2DESs
\cite{vakili04}. Furthermore, electrons in the Z valley are very
similar to electrons in Si metal-oxide-semiconductor field-effect
transistors (MOSFETs) (e.g. similar band effective mass, Land\'{e}
g-factor, etc.) except for the lack of valley degeneracy, thereby
providing a control system for the study of effects related to
that degeneracy. Though there are previous reports of occupation
of the Z valley in AlAs \cite{vankesteren89, vandestadt96}, these
early samples had low mobilities ($\sim$ 0.37 m$^2$/Vs at 0.5 K),
and subsequent experiments in AlAs have almost exclusively
involved X and Y valley occupation.

\begin{figure}
\centering
\includegraphics[scale=0.34]{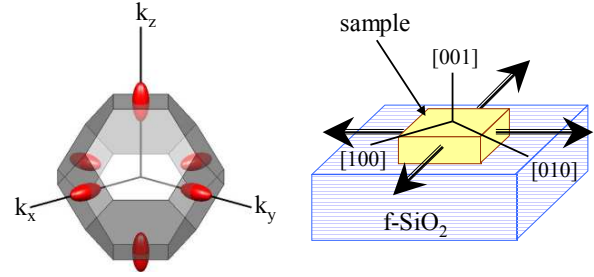}
\caption{Left: the Fermi surface of bulk AlAs and the first
Brillouin zone. Right: schematic illustration of a sample glued on
fused SiO$_2$, with the arrows indicating the direction of the
resulting thermally induced strain.}
\end{figure}

We have studied several different AlAs QWs with thicknesses less
than or close to {\it t}$_{c}$.  All samples were patterned in an
L-shaped Hall bar configuration with arms parallel to the in-plane
crystal axes. Ohmic contacts were made by depositing AuGeNi and
alloying in a reducing environment.  Metallic front and back gates
allow for {\it in situ} control of the electron density, {\it n},
and facilitate populating the QW via a recently developed
field-effect persistent photoconductivity (PPC) technique
\cite{depoortere03}. We made measurements in a pumped $^3$He
refrigerator with a base temperature of 0.3 K and employed
standard low-frequency lock-in techniques.

The layer structure near the surface of the wafer containing our
thinnest QW is similar to that in Ref. \cite{depoortere02} except
for a narrower QW width. A 45 $\AA$ AlAs QW is grown atop a thick
layer of Al$_{0.39}$Ga$_{0.61}$As and separated from a Si
$\delta$-doping layer by a 750 $\AA$ spacer also made of
Al$_{0.39}$Ga$_{0.61}$As.  This front-side modulation-doping,
together with the aforementioned PPC technique, considerably
improves the mobility in thick AlAs QWs with the (X,Y) valleys
occupied \cite{depoortere02}. The QW layer is flanked by thin
layers of pure GaAs, separating it from the AlGaAs barriers to
reduce interface roughness at the QW edges and limit the
penetration of the electronic wavefunction into the AlGaAs where
alloy scattering can take place.

\begin{figure}
\centering
\includegraphics[scale=0.34]{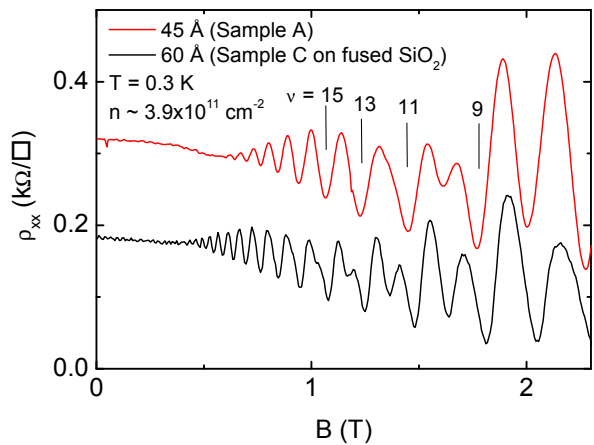}
\caption{$\rho_{xx}$ vs. {\it B} for 2DESs of approximately equal
density confined to two AlAs QWs of different width.  The sample
containing the 60 $\AA$ QW is glued to a fused silica block.}
\end{figure}

A typical magnetoresistivity ($\rho_{xx}$) trace for a sample (A)
from this wafer is shown in Fig. 2.  Shubnikov-de Haas (SdH)
oscillations are well developed and persist to fields as low as
{\it B} = 0.6 T and filling factors as high as $\nu$ = 25. This is
a considerable improvement over the only other transport
measurement for Z valley electrons in AlAs of which we are aware
\cite{vandestadt96}, where SdH oscillations first become
perceptible at {\it B} = 5 T.  We have confirmed that the Z valley
is occupied in our samples by extracting the effective mass from
the temperature dependence of SdH oscillation amplitudes at high
{\it n} \cite{vakili04}.

The dependence of mobility, $\mu$, on {\it n} for sample A is
exhibited in Fig. 3 (triangles). A peak mobility of 5.1 m$^2$/Vs
is achieved near {\it n} = 3.5 x 10$^{11}$ cm$^{-2}$, which is
more than an order of magnitude higher than the peak $\mu$
reported for the Z valley sample in Ref. \cite{vandestadt96} and
comparable to the highest mobility achieved in Si-MOSFETs. Though
this is a dramatic improvement, it is still about six times lower
than the peak $\mu$ values reported for thicker AlAs QWs with
(X,Y) valley occupation \cite{depoortere02}, despite essentially
identical growth conditions. Additionally, there is a saturation
and even slight decline of $\mu$ as {\it n} increases past the
peak value. Both facts are consistent with the hypothesis that
$\mu$ is limited by interface-roughness scattering (IRS) at high
{\it n} in the AlAs QWs that are thin enough to allow Z valley
occupation.

Since the effects of IRS rapidly weaken with increasing QW
thickness \cite{sakaki87}, we measured a second sample (B) that
contains a slightly thicker AlAs QW (50 $\AA$) but is otherwise
identical to A. $\mu$ vs. {\it n} for this QW is shown in Fig. 3
(circles). Indeed, $\mu$ attains a peak value of 5.5 m$^2$/Vs,
higher than sample A, but then drops rapidly for higher {\it n}.
This drop begins near the same {\it n} where we observe Landau
level crossings in $\rho_{xx}$ that we can attribute to the (X,Y)
valley(s).  The rapid drop in $\mu$, therefore, is likely
associated with inter-valley scattering at the onset of (X,Y)
valley population.

\begin{figure}
\centering
\includegraphics[scale=0.34]{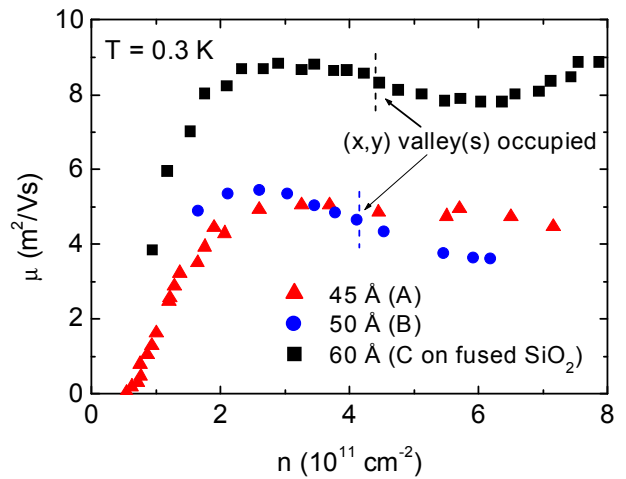}
\caption{$\mu$ vs. {\it n} for 2DESs in QWs with three different
widths. Sample C is glued to a fused silica block.  The
approximate densities at which additional valleys become occupied
in samples B and C are indicated.}
\end{figure}

In order to side-step this mobility limiting dilemma, we devised a
technique that allows preferential population of the Z valley in
QWs thicker than {\it t}$_{c}$ (see Fig. 1).  First, after the
conventional processing described above (patterning, formation of
the contacts, and front gate deposition) the sample is thinned to
$\simeq$ 150 $\mu$m. We then deposit a back gate and glue the
sample to a smooth block of fused silica (f-SiO$_2$) using a
two-component epoxy and curing technique described previously
\cite{shayegan03}. We used f-SiO$_2$ blocks cut from
optical-quality mirrors with their metallization removed to ensure
smoothness of the surface on which the sample is glued. The
purpose of gluing the samples to f-SiO$_2$ is related to this
material's low thermal expansion coefficient
($\alpha_{f-SiO_{2}}$). As an ordinary sample is cooled, it
contracts along all three directions in proportion to the thermal
expansion coefficient of GaAs, $\alpha_{GaAs}$, which is 6.4 x
10$^{-6}$/K at room temperature ({\it T}$_{r}$) \cite{feder68}.
When thinned and affixed to f-SiO$_{2}$, however, the sample's
contraction follows the f-SiO$_2$ along the in-plane axes. Since
$\alpha_{f-SiO_2}$ (= 0.55 x 10$^{-6}$/K at {\it T}$_{r}$
\cite{okaji95}) is lower than $\alpha_{GaAs}$, this causes a
temperature induced, biaxial, tensile strain in the plane of the
QW and a compressive strain in the [001] direction along which the
sample is free (Fig. 1).  As a result, the (X,Y) valleys are
raised in energy with respect to the Z valley.

To acquire an estimate of this energy and to further elucidate the
technique, we consider the plane stress boundary conditions.  We
assume that the relevant Poisson ratios and Young's moduli are
temperature independent.  Since the temperature at which our
experiments are performed is low (0.3 K), we take it to be zero.
The strain along each direction is:

\begin{equation}
\epsilon_{xx} = (\sigma_{xx} - \upsilon\sigma_{yy})/E +
\int_{0}^{T_{r}}\alpha_{GaAs}dT
\end{equation}
\begin{equation}
\epsilon_{yy} = (\sigma_{yy} - \upsilon\sigma_{xx})/E +
\int_{0}^{T_{r}}\alpha_{GaAs}dT
\end{equation}
\begin{equation}
\epsilon_{zz} = - \upsilon(\sigma_{xx}/E + \sigma_{yy}/E) +
\int_{0}^{T_{r}}\alpha_{GaAs}dT
\end{equation}
where {\it E} is the Young's modulus of GaAs along the crystal
axes, $\upsilon$ = 0.32 is the Poisson ratio, $\sigma$ is stress,
$\epsilon$ is strain, and we have used the free boundary condition
in the z direction, $\sigma_{zz}$ = 0. We have ignored edge
effects since our active area is far from the sample edges.
Setting $\epsilon_{xx}$ = $\epsilon_{yy}$ =
$\int_{0}^{T_{r}}\alpha_{f-SiO_{2}}dT$ and solving for the net
strain between the Z and (X,Y) directions gives,

\begin{equation}
\epsilon_{net} = \epsilon_{zz} - \epsilon_{xx} = \kappa
\int_{0}^{T_{r}}(\alpha_{GaAs} - \alpha_{f-SiO_{2}})dT
\end{equation}
with $\kappa$ = (1+$\upsilon$)/(1-$\upsilon$).  Utilizing the
temperature dependent $\alpha_{GaAs}$ \cite{feder68} and
$\alpha_{f-SiO_{2}}$ \cite{okaji95}, and the X-point deformation
potential of AlAs, {\it E}$_2$ = 5.8 eV \cite{charbonneau91}, we
can calculate the valley splitting associated with the thermally
induced strain: {\it E}$_{net}$ = {\it E}$_{2}\epsilon_{net}$
$\simeq$ 9 meV. Subtracting this from the 23 meV of lattice
mismatch induced valley splitting, we acquire the final
strain-induced splitting between Z and (X,Y) valleys. This would
yield an upward shift of {\it t}$_c$ by approximately 20 $\AA$,
based on the confinement energy calculations of Ref.
\cite{vankesteren89}.

We have tested this technique on a third sample (C), a 60 $\AA$ QW
that is otherwise structurally identical to the other two samples,
and have indeed observed occupation of only the Z valley at low
{\it n}. Control samples taken from nearby sample C on the same
wafer but not affixed to f-SiO$_2$ show only (X,Y) valley
occupation, with no indication of the Z valley even at the highest
attainable {\it n}.  This implies that the thermally induced
strain in the sample glued to f-SiO$_2$ is forcing the Z valley
occupation, as expected.  The $\mu$ limiting effects of IRS appear
to be significantly reduced in the 60 $\AA$ QW; as shown in Fig. 3
(squares), a peak $\mu$ of 8.8 m$^2$/Vs is achieved. The improved
quality is also evident in $\rho_{xx}$ (Fig. 2), with SdH
oscillations visible down to {\it B} = 0.4 T ($\nu$ = 37), and
spin splitting resolved at a 30 $\%$ lower field than in sample A.
At higher {\it n}, there is evidence for (X,Y) valley occupation,
and consequently $\mu$ begins to drop once again.

Although our experimental results are in qualitative agreement
with the simple calculation given above, there is one complicating
factor that we have neglected. The epoxy we have used to glue our
samples to the f-SiO$_{2}$ is known to creep at higher
temperatures, and this effect is believed to persist down to about
200 K \cite{shkolnikov05}.  This will reduce somewhat the
effectiveness of our technique, as the strain will not be fully
transmitted over the entire temperature range.  It may be
desirable to use a glue that transmits strain more effectively at
higher temperatures, or even to glue the sample to any of the
novel materials that exhibit negative coefficients of thermal
expansion, though there may eventually be a risk of cracking the
sample for too large of a strain.

We thank the NSF and ARO for financial support.

\break

\end{document}